\documentclass{aa}  

\usepackage{graphicx}
\usepackage{txfonts}
\usepackage{color}
\usepackage{bm}
\usepackage{media9}
\begin{document} 

   \title{ COCONUT: Toward practical time-evolving Sun-to-Earth magnetohydrodynamic modeling }
   \author{H. P. Wang \inst{1}
          \and
          S. Poedts\inst{1,2}
          \and
          A. Lani\inst{1}
          \and 
          R. Dhib\inst{1}
          \and
          L. Linan\inst{1}
          \and
          T. Baratashvili\inst{1}
          \and
          F. Zhang\inst{3,4}
          \and
          Q. Noraz\inst{1,4}
          \and
          H.-J. Jeong\inst{1,5}
          \and
          N. Wijsen\inst{1}
          \and
          M. Condoluci\inst{1}
          \and
          L. Y. Dong\inst{1}
          \and
          J. Y. Liu\inst{1}
          \and
          R. Zhuo\inst{1}
          \and
          M. Najafi-Ziyazi\inst{1}
          \and
          K. Arabuli\inst{1}
          \and
          M. Flossie\inst{1}
          \and
          J. Magdalenic\inst{1,6}
          \and
          B. Schmieder\inst{1,7,8}
          }
   \institute{Centre for Mathematical Plasma-Astrophysics, Department of Mathematics, KU Leuven, Celestijnenlaan 200B,
      3001 Leuven, Belgium\\
  \email{Stefaan.Poedts@kuleuven.be}\\
  \email{andrea.lani@kuleuven.be}\\
  \email{haopeng.wang1@kuleuven.be}
\and 
Institute of Physics, University of Maria Curie-Skłodowska, ul. Radziszewskiego 10, 20-031 Lublin, Poland 
\and
Institute of Theoretical Astrophysics, University of Oslo, PO Box 1029 Blindern, 0315 Oslo, Norway
\and
Rosseland Centre for Solar Physics, University of Oslo, PO Box 1029 Blindern, 0315 Oslo, Norway
\and
School of Space Research, Kyung Hee University, Yongin, 17104, Republic of Korea
\and
Solar-Terrestrial Centre of Excellence—SIDC, Royal Observatory of Belgium, 1180 Brussels, Belgium
\and
Observatoire de Paris, LIRA, UMR8254 (CNRS), F-92195 Meudon Principal Cedex, France
\and
LUNEX EMMESI COSPAR-PEX Eurospacehub, Kapteyn straat 1, Noordwijk 2201 BB, Netherlands}
  \abstract
   {To enable timely protective actions against severe space weather, it is essential to develop advanced Sun-to-Earth magnetohydrodynamic (MHD) models to deliver timely, high-fidelity, and comprehensive space weather forecasts. 
   Due to computational efficiency and numerical stability limitations, coronal simulations constrained by static magnetograms are typically performed first and then used to drive inner-heliosphere (IH) models, assuming that the coupling has a negligible influence on the simulation results and that variations in the background coronal structures are minor over the simulation period. 
   } 
    { In this paper, we calculate the Sun‑to‑Earth coronal and wind evolutions using a single time-evolving MHD model, showing that implicit MHD models have the potential to meaningfully simplify and improve the overall Sun‑to‑Earth modelling pipeline.
    }
   { We extend the implicit time‑evolving coronal MHD model COCONUT out to 1 AU, and utilise it to investigate solar coronal and wind evolutions around a solar maximum Carrington rotation (CR). 
   We compare quasi-steady-state and time-evolving Sun-to-Earth simulations to evaluate the impact of the inner-boundary magnetic field evolution, which is neglected in steady-state simulations, on background plasma parameters.
   Comparisons with commonly used coupled Sun-to-Earth simulations are also conducted to further validate and assess the Sun-to-Earth model COCONUT.
   }
   { The results show that the time-evolving implicit MHD modelling approach yields noticeable differences compared to oversimplified steady-state simulations, and is efficient enough for practical applications, such as faster-than-real-time daily space weather forecasting. Modelling the solar corona and wind using a single MHD model simplifies the modelling pipeline and avoids uncertainties associated with coupling different coronal and IH models. The noticeable differences in the temporal evolution of plasma parameters at the L1 and L5 points highlight the need to use continuously evolving, synchronised magnetic field observations to improve global coronal and solar wind simulations, whereas the overall consistent evolutionary trend reveals the reliability of using L5 observations to forecast solar wind conditions near Earth about four days in advance.}
   {}

   \maketitle
%
\section{Introduction}
The solar corona and wind consist of highly ionised, magnetised plasma that fills the entire Sun-to-Earth space. The Sun-to-Earth system involves physical processes that span a wide range of temporal and spatial scales, with the coronal component being particularly complex and challenging for numerical simulations. Magnetohydrodynamic (MHD) processes dominate various coronal dynamics \citep{Cranmer_2005}, including solar flares, coronal mass ejections (CMEs), plasma jets, coronal waves, coronal heating, and magnetic reconnection \citep{cox1991solar,Aschwanden2005,bookPriest_2014}. These coronal activities continuously shape the heliospheric environment and can trigger hazardous space weather events, which can damage satellite electronics, disrupt navigation and power systems, corrode pipelines, and pose risks to human health \citep{Eastwood2017,Feng2020book}. 
There is an urgent need to develop advanced Sun-to-Earth model chains \cite[e.g.][]{Groth2000,Feng_2011Chinese,Feng_2013Chinese,Feng2020book,Jin_2012,TOTH2012870,Owens2017,Pomoell2018020,Hoeksema2020,Poedts_2020,Gombosi2021,Hayashi_2021,Yang_2021,Linker2024,Wraback_2025} that effectively balance computational efficiency, numerical stability, and physical fidelity, in order to better understand space weather mechanisms and ultimately provide reliable space weather forecasts hours to days in advance.

\begin{figure*}[htpb]
\begin{center}
  \vspace*{0.01\textwidth}
    \includegraphics[width=0.9\linewidth,trim=1 1 1 1, clip]{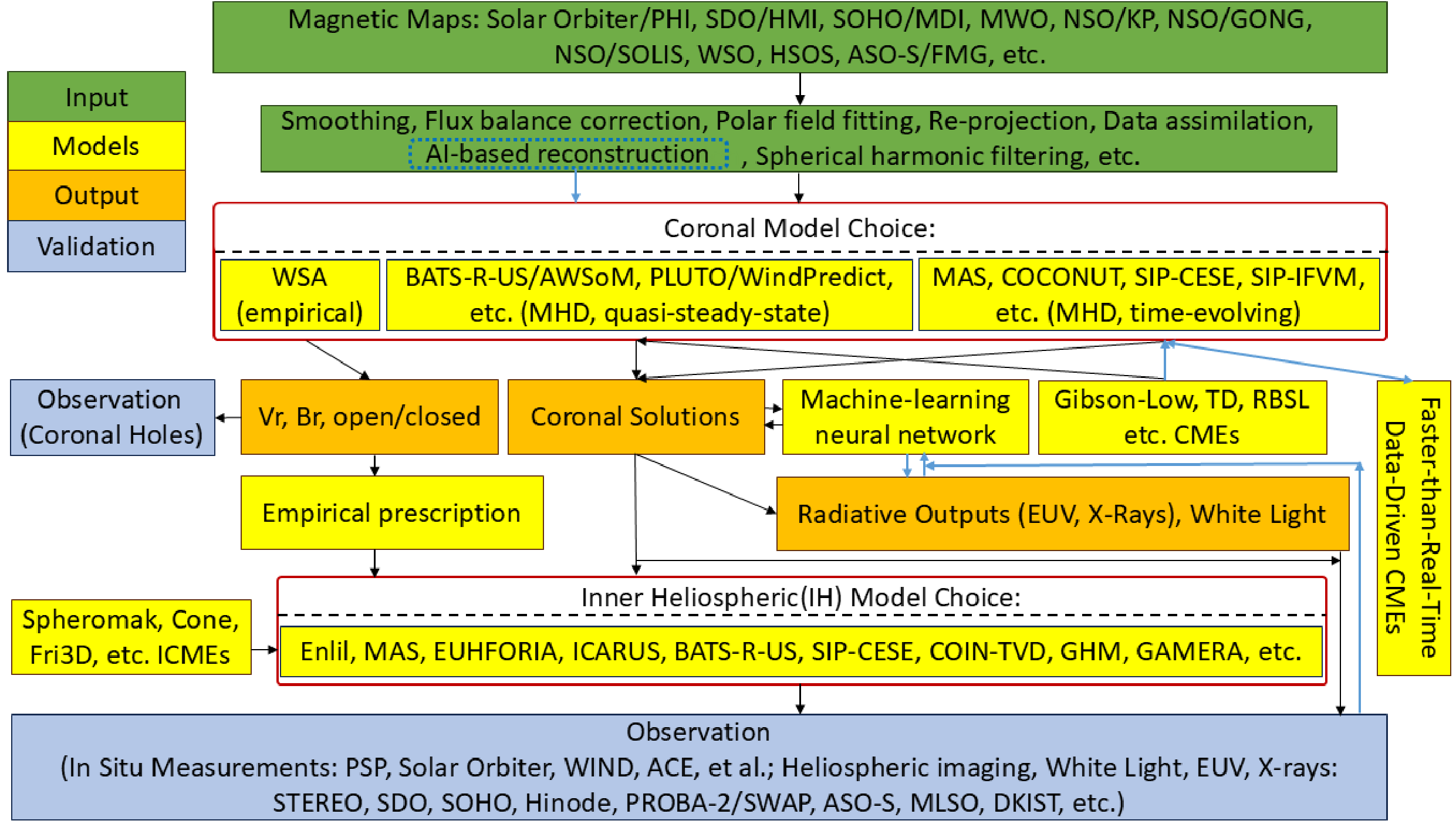}
\end{center}
\caption{Illustration of the Sun-to-Earth model chain; the blue lines highlight the modules under development.}\label{SuntoEarthModelChain}
\end{figure*}
In the Sun-to-Earth model chain, as illustrated in Fig.~\ref{SuntoEarthModelChain}, the synoptic magnetograms derived from photospheric observations provide the inner-boundary magnetic field constraints for the coronal model, and the coronal simulation results in the super-Alfv{\'e}nic and supersonic regions provide the inner-boundary conditions for the inner heliospheric (IH) model, which further provides the boundary conditions for the geomagnetic model. Due to complex coronal dynamics and low-$\beta$ conditions \cite[$\beta$ is the ratio of thermal to magnetic pressure;][]{Feng_2021,Hayashi_2021,Brchnelova2025,wang2025sipifvmobservationbasedmagnetohydrodynamicmodel,wang2025sipifvmtimeevolvingcoronalmodel,Wang_COCONUT_DecE} around active regions near the solar surface \citep{Bourdin2017}, coronal MHD modelling usually suffers from limited computational efficiency and insufficient numerical stability, particularly around solar maximum. As a compromise, some empirical coronal models \citep{Arge2003ImprovedMF,Yangzicai2018} are adopted \citep{Merkin2016,Pomoell2018020,Shen_2018,Perri_2023}, at the expense of physical accuracy and completeness \citep{Samara_2021}. 
Additionally, the time steps of widely used explicit and semi-implicit coronal MHD models are seriously limited by numerical stability \citep{Lionello1999346} to a few seconds or less \citep{ManchesterIV2004,Feng_2012,FengandLiu2019,Caplan_2017,Caplan_2019}, whereas time steps in IH MHD models can reach ten minutes or more \citep{Detman2006,Hayashi2012}, depending on the grid resolution. 
Consequently, using coronal simulation results at a prescribed interface as inner-boundary conditions to drive the IH model \citep{van_der_Holst_2010,van_der_Holst_2022,TOTH2012870,Torok_2018,Poedts_2020,Baratashvili2024,Linan_2025} is less computationally intense than employing a single model, constrained by the smallest local time step in the whole computational domain, to simulate both the coronal and IH regions simultaneously \citep{Lionello_2013}. 

Due to differences in numerical schemes, grid structures, boundary condition prescriptions, and physical assumptions, coupling two models covering different computational domains via a prescribed interface may introduce artifacts near the interface \citep{Riley2021}. A coronal simulation extending from the solar surface to 50~$R_s$, constrained by a static magnetogram and including an inserted flux rope to initiate a CME, was conducted and then a series of solutions extracted at 20~$R_s$ from this simulation was used to drive an IH model spanning 20-50~$R_s$ \citep{Lionello_2013}.
It shows that the latter case can produce IH solutions that are visually indistinguishable from those of the former case, with discrepancies in most simulated solutions not exceeding $1\%$.
However, it remains unclear to what extent IH simulation results can be influenced by different choice of IH models, particularly in time-evolving simulations driven by time-dependent magnetograms. This paper addresses this issue by conducting a comparison between simulation results obtained from a single time-evolving Sun-to-Earth model and those from an IH model driven by the outputs of the Sun-to-Earth model at 0.1 AU.

In general, MHD models of the solar corona and wind can be classified into two categories: quasi-steady models and time-evolving models. Models constrained by a single static magnetogram are referred to as quasi-steady-state models, whereas those driven by a sequence of time-evolving magnetograms are defined as time-evolving models. The former assume a (quasi-) steady corona during one CR, while the later are time-accurate and driven by continuously updated magnetograms, enabling a more realistic representation of the evolving solar coronal and wind structures and laying the foundation for more accurate solar wind and CME simulations \citep{Lionello_2023}. A more detailed description of these two classes of coronal MHD models can be found in \cite{Feng2020book}, \cite{Wang2025_FirsttimeevolvingCOCONUT,wang2025COCONUTMayEvent,wang2025sipifvmtimeevolvingcoronalmodel,Wang_COCONUT_DecE} and \cite{Downs2025PSI}. Among these models, the time-evolving coronal models COolfluid COroNal UnsTructured \cite[COCONUT;][]{Wang2025_FirsttimeevolvingCOCONUT,wang2025COCONUTMayEvent,Wang_COCONUT_DecE} and Solar Interplanetary Phenomena Implicit Finite Volume Method \cite[SIP-IFVM;][]{wang2025sipifvmtimeevolvingcoronalmodel} adopt novel implicit algorithms that significantly increase computational efficiency by adopting a large time step, demonstrating the strong potential of implicit approaches for faster-than-real-time coronal evolution simulations in practical applications. 

With advances in implicit algorithms \cite[e.g.][]{TOTH2012870,WANG201967,Feng_2021,Perri_2022,Perri_2023,Wang_2022,Wang2022_CJG,WangSIPtheoriticalCME,wang2025sipifvmobservationbasedmagnetohydrodynamicmodel}, as well as improvement in computational capabilities such as GPU acceleration \cite[e.g.][]{Feng_2013,Caplan_2019,WANG2019181}, it has become possible to simulate the entire Sun-to-Earth domain efficiently using a single implicit MHD model, which simplifies the modelling pipeline and avoids uncertainties associated with coupling separate models. Additionally, the local time stepping approach \citep{Groth2000,TOTH2012870}, which employs different time steps in different grid cells based on local numerical stability constraints, can be an option to further accelerate the computational efficiency of MHD solar coronal and wind models. 
In this paper, we extend the implicit time-evolving coronal model COCONUT to beyond 1 AU, thereby establishing the first fully implicit 3D global time-evolving Sun-to-Earth MHD model. We employ it to simulate solar coronal and wind evolutions during solar maximum and minimum CRs to evaluate evolutions of plasma parameters at the L1 and L5 points, providing a preliminary investigation for future multi-satellite solar observations. 

COCONUT is a recently developed implicit MHD coronal model built on the Computational Object-Oriented Libraries for Fluid Dynamics (COOLFluiD) framework \citep{kimpe2,lani1,lani13}\footnote{\url{https://github.com/andrealani/COOLFluiD/wiki}}. In quasi-steady-state coronal MHD simulations, COCONUT achieved a speed up of 35$\times$ \citep{Perri_2022,Perri_2023} compared with WindPredict \citep{Perri2018SimulationsOS,Parenti_2022}, an explicit coronal model based on the PLUTO code \citep{Mignone_2007}. COCONUT has also been developed into an efficient time-evolving coronal MHD model capable of simulating continuous coronal evolutions over an entire solar minimum CR on approximately 1.5 million (M) grid cells by only 10 thousand CPU hours \citep{Wang2025_FirsttimeevolvingCOCONUT}. By employing a positivity-preserving (PP) approach for density and thermal pressure and limiting the inner-boundary magnetic field to not significantly exceed 30 Gauss (G) via a spherical harmonic filtering \citep{MCCLARREN20105597, Murteira2025}, the time-evolving COCONUT model is capable of effectively simulating coronal evolution during solar maximum \citep{wang2025COCONUTMayEvent}. 
Furthermore, the decomposed energy strategy, combined with a Harten-Lax-van Leer Riemann solver \citep{Harten1983,Feng2020book,Wang_2022} with an appropriate dissipation term incorporated into the energy equation component in low-$\beta$ regions, enables COCONUT to simulate coronal evolutions during solar maximum even with magnetic field strengths reaching 100 G and with plasma $\beta$ below $10^{-3}$ in the low corona near active regions \citep{Wang_COCONUT_DecE}.  
This paper further demonstrates that a single MHD model employing time-accurate implicit temporal integrations can efficiently simulate the Sun-to-Earth coronal and wind evolutions with the required accuracy.

Based on the above considerations, the paper is organised as follows. In Section \ref{NumericalAlgorithm}, we introduce the numerical algorithm used to perform the time-evolving Sun-to-Earth coronal and wind simulations. In Section~\ref{sec:Numerical Results}, we present simulation results around CR 2282.
Steady-state and time-evolving simulation results of the Sun-to-Earth model COCONUT are compared. Variations of plasma parameters at the L1 and L5 points calculated by the Sun-to-Earth model COCONUT and the coupled coronal and IH models are compared. 
In Section~\ref{sec:Conclusion}, we summarise the features of the implicit time-evolving Sun-to-Earth MHD modelling approach and provide concluding remarks. 

\section{Numerical algorithm}\label{NumericalAlgorithm}
The governing equations and the discretization methods are essentially the same as those in \cite{Wang_COCONUT_DecE}. The main differences are as follows. The simulations in this paper are conducted in a co-rotating Carrington heliographic coordinate system \citep{Burlaga1984MHDPI,FRANZ2002217}. A Venkatakrishnan limiter \citep{VENKATAKRISHNAN1993,Wang2025_FirsttimeevolvingCOCONUT} is implemented to the velocity where the solar wind speed exceeds a reference Alfv{\'e}nic velocity, $V_{A,\rm ref}=480.25~\rm Km\cdot S^{-1}$, derived from a reference magnetic field strength of $2.2\times 10^{-4}~\rm Tesla$ and a reference plasma density of $1.67\times 10^{-13}~\rm Kg \cdot m^{-3}$. Additionally, the computational domain is extended and the grid mesh is modified to cover the entire Sun-to-Earth region, and a PP measure is applied to the temperature.

\subsection{The governing equations and grid system}\label{Governingequations} 
The governing equation adopted in this paper is the same as in \cite{Wang_COCONUT_DecE}, and can be described in the following form:
\begin{equation}\label{MHDinsolarwind}
\frac{\partial \mathbf{U}}{\partial t}+\nabla \cdot \mathbf{F}\left(\mathbf{U}\right)=\mathbf{S}\left(\mathbf{U},\nabla \mathbf{U}\right) \,.
\end{equation}
In Eq.~\ref{MHDinsolarwind}, $t$ refers to time, $\mathbf{U}=\left(\rho, \rho \mathbf{v}, \mathbf{B}, E_1, \psi\right)^T$ means the vector of conservative variables, $\nabla \mathbf{U}$ and $\mathbf{F}\left(\mathbf{U}\right)$ denote the spatial derivative of $\mathbf{U}$ and the inviscid flux vector. Here $\rho$ is the plasma density, $\mathbf{v}=\left(u,v,w\right)$ and $\mathbf{B}=\left(B_x,B_y,B_z\right)$ means the velocity and magnetic field in the Cartesian coordinate system, $E_1=\frac{p}{\gamma-1}+\frac{1}{2}\rho\left|\mathbf{v}\right|^{2}$ is the decomposed energy density \citep{Wang_COCONUT_DecE} with $\gamma=\frac{5}{3}$ being the adiabatic index, and $\psi$ demotes the Lagrange multiplier defined in the hyperbolic generalised Lagrange multiplier method used for constraining the divergence errors \citep{Dedner2002JCoPh,YALIM20116136}. Additionally, $\mathbf{S}\left(\mathbf{U},\nabla \mathbf{U}\right)=\mathbf{S}_{\rm gra}+\mathbf{S}_{\rm heat}+\mathbf{S}_{\rm{DECOMP}}+\mathbf{S}_{\rm Cori}$, with $\mathbf{S}_{\rm gra}$ and $\mathbf{S}_{\rm DECOMP}$ denoting vectors of the source terms correspond to the gravitational force and the source term derived from the decomposed energy equation, as described in \cite{Wang_COCONUT_DecE}. The same optically thin radiative loss $Q_{rad}$ and the Spitzer or collisionless thermal conduction term $-\nabla \cdot \mathbf{q}$ as in \cite{Wang_COCONUT_DecE} are considered in the heating source terms $\mathbf{S}_{\rm heat}=-\nabla \cdot \mathbf{q}+Q_{rad}+Q_{H}$, and the coronal heating $Q_{H}$ is empirically defined as follows \citep{Mok_2005,Downs2010,Baratashvili2024,WangSIPtheoriticalCME,wang2025COCONUTMayEvent}:
$$
Q_{H}=H_0 \cdot \left|\mathbf{B}\right| \cdot e^{-\frac{r-R_s}{\lambda}}\,,
$$
where $\lambda = 0.7\,R_s$ and $H_0 = 2 \cdot 10^{-2}\,{\rm J\,m^{-3}\,s^{-1}\,T^{-1}}$.
In this paper, the simulations are conducted in a co-rotating coordinate system; therefore, source terms associated with the Coriolis force and centrifugal acceleration are also included and computed as follows:
$$
\mathbf{S}_{\rm Cori}=-\rho
                        \begin{pmatrix}
                        0\\
                        \mathbf{\omega} \times \left(\mathbf{\omega} \times \mathbf{r}\right)+2\mathbf{\omega} \times \mathbf{v} \\
                        \mathbf{0}\\                \mathbf{v}\cdot\left(\mathbf{\omega} \times \left(\mathbf{\omega} \times \mathbf{r}\right)\right)\\
                        0
                        \end{pmatrix}\,,
$$
where $|\mathbf{\omega}|=2.97\times10^{-6} ~ \rm radian ~ second^{-1}$ denotes the sidereal angular speed of the Sun and $\mathbf{r}$ is the position vector.

The computational domain for the Sun-to-Earth simulation is a spherical shell extending from 1.01 to approximately 256~$R_s$ that is discretised using an unstructured sixth-level subdivided geodesic meshes \citep{Brchnelova2022,Wang2025_FirsttimeevolvingCOCONUT,wang2025COCONUTMayEvent}. It consists of 185 radial layers of gradually expanding truncated pentagonal-pyramidal cells, with 20,480 cells in each layer. For the coronal simulations, the computational domain extends from 1.01 to about 25~$R_s$, comprising 117 radial layers of grid cells which are identical to those in the corresponding domain of the Sun-to-Earth grid mesh.

\subsection{Boundary-condition setups}\label{BCinner}
In this paper, we first perform a quasi-steady-state coronal or Sun-to-Earth simulation constrained by a static magnetogram on March 12, 2024. We then use cubic Hermite interpolation to update the inner-boundary magnetic field at each physical time step, based on a series of hourly-updated GONG-zqs photospheric magnetograms\footnote{\url{https://gong.nso.edu/data/magmap/QR/zqs/202403/}} \citep{LiHuichao2021,Perri_2023} that have been extrapolated to the low corona by a 50th-order spherical-harmonic potential field solver with a filter \citep{MCCLARREN20105597, wang2025COCONUTMayEvent, Murteira2025}, as described in \cite{wang2026mhdmodellingopenflux}, to drive the subsequent solar coronal and wind evolutions during the following simulation periods \citep{Wang2025_FirsttimeevolvingCOCONUT,Wang_COCONUT_DecE}. 
Meanwhile, the inner-boundary thermal pressure and plasma density are set to $0.0042~\rm Pa$ and $1.67 \times 10^{-13}~\rm kg,m^{-3}$, respectively, with the inner-boundary temperature being $1.5 \times 10^6~\rm K$.  
Additionally, the inner-boundary density is empirically adjusted to be nonhomogeneous \citep{wang2026mhdmodellingopenflux} based on the surface-to-corona simulation results performed with the \textit{Bifrost} radiative MHD code (Noraz et al. 2026b), and the velocity vector, the tangential magnetic field at the inner boundary, and the outer boundary conditions are treated the same as in \cite{Wang2025_FirsttimeevolvingCOCONUT}.

\subsection{PP measure for temperature}\label{PP_T}
During numerical calculations, the thermal pressure is updated as $p=\left(\gamma-1\right)\cdot\left(E_1-\frac{1}{2}\rho\left|\mathbf{v}\right|^{2}\right)$. Meanwhile, the temperature of the bulk plasma is derived from the equation of state $T=\frac{p}{\Re~\rho}$ with $\Re$ denoting the gas constant \citep{Wang2025_FirsttimeevolvingCOCONUT}. Here $\Re=\frac{2*k_B}{m_{\rm cor}*m_H}$ with $k_B=1.3806503 \times 10^{-23}\rm ~J~K^{-1}$ denoting the Boltzmann constant, $m_{\rm cor}=1.27$ is the molecular weight and $m_H=1.67262158 \times 10^{-27}~ \rm kg$ represents the mass of hydrogen.
The thermal pressure of the bulk plasma in the IH component is extremely small, therefore the discretisation errors in $E_1$ and $\rho\left|\mathbf{v}\right|^{2}$ can sometimes lead to a negative thermal pressure, resulting in a nonphysical negative temperature. To prevent such issues, we apply a PP measure to the temperature as follows:
\begin{equation}\label{T_PP}
p=\Upsilon\cdot T_{\min}\cdot \Re\cdot \rho+\left(1-\Upsilon\right)\cdot p_{o} 
\end{equation}
where $T_{\rm min}=10^4$~K, $\Upsilon=0.5+0.5\cdot \tanh\left(\frac{T_{\min}-T_o}{20.0}\cdot \pi\right)$, $p_o$ and $T_o$ are the originally updated thermal pressure and temperature.

\section{Numerical results}\label{sec:Numerical Results}
In this section, we compare the steady-state and time-evolving Sun-to-Earth simulation results. 
We present the plasma parameters at the L1 and L5 points simulated by the time-evolving Sun-to-Earth MHD model COCONUT, together with those obtained from the IH model EUHFORIA \citep{Pomoell2018020,Poedts_2020,Linan_2025} driven by the the time-evolving Sun-to-Earth COCONUT simulation results at 0.1~AU. 
Additionally, the time-evolving Sun-to-Earth COCONUT simulation results at 0.1~AU and 3~$R_s$ are compared with the coronal COCONUT simulation results in Appendix~\ref{IHimpactonCoronal} to assess the influence of the inner heliosphere component on coronal simulations.

The time-evolving simulations are driven by about 740 hourly updated magnetograms spanning from 22:14 on March 12, 2024, to 22:14 on April 12, 2024.
Executed on 1,440 CPUs of the WICE cluster, part of the Tier-2 supercomputer at the Vlaams Supercomputer Centrum\footnote{\url{https://www.vscentrum.be/}}
, the Sun-to-Earth simulation, with a time step of 5 minutes, achieves a 22.5$\times$ speedup over real-time solar coronal and wind evolution.

\subsection{Time-evolving versus quasi-steady-state Sun-to-Earth simulations}\label{sec:TEversusQSS}
In this subsection, we compare the time-evolving simulation results with quasi-steady-state simulation results constrained by the magnetogram corresponding to the initial time of the time-evolving simulation. The heliolongitude $\phi$ in the quasi-steady state coronal simulation is mapped to the corresponding time $t$ in the time-evolving simulations as follows \citep{Wang2025_FirsttimeevolvingCOCONUT}: 
\begin{equation}\label{phitotime}
\phi=\phi_0-360^{\circ}\times\frac{t-t_0}{T_{\rm CR}}\,,
\end{equation}
where $t_0$ denotes the corresponding time of the magnetogram used in the quasi-steady-state simulation, $\phi_0$ is the selected heliolongitude at $t_0$ in the time-evolving simulation, and $T_{\rm CR}$ represents the duration of the corresponding CR. In this research work, $\left(\phi_0,~t_0\right)$= $\left(0^{\circ},~1~{\rm hour}\right)$ and $\left(300^{\circ},~1~{\rm hour}\right)$ are selected for the L1 and L5 points, respectively, and $T_{\rm CR}=654.5$ hours.

\begin{figure*}[htpb]
\begin{center}
  \vspace*{0.01\textwidth}
    \includegraphics[width=0.8\linewidth,trim=1 1 1 1, clip]{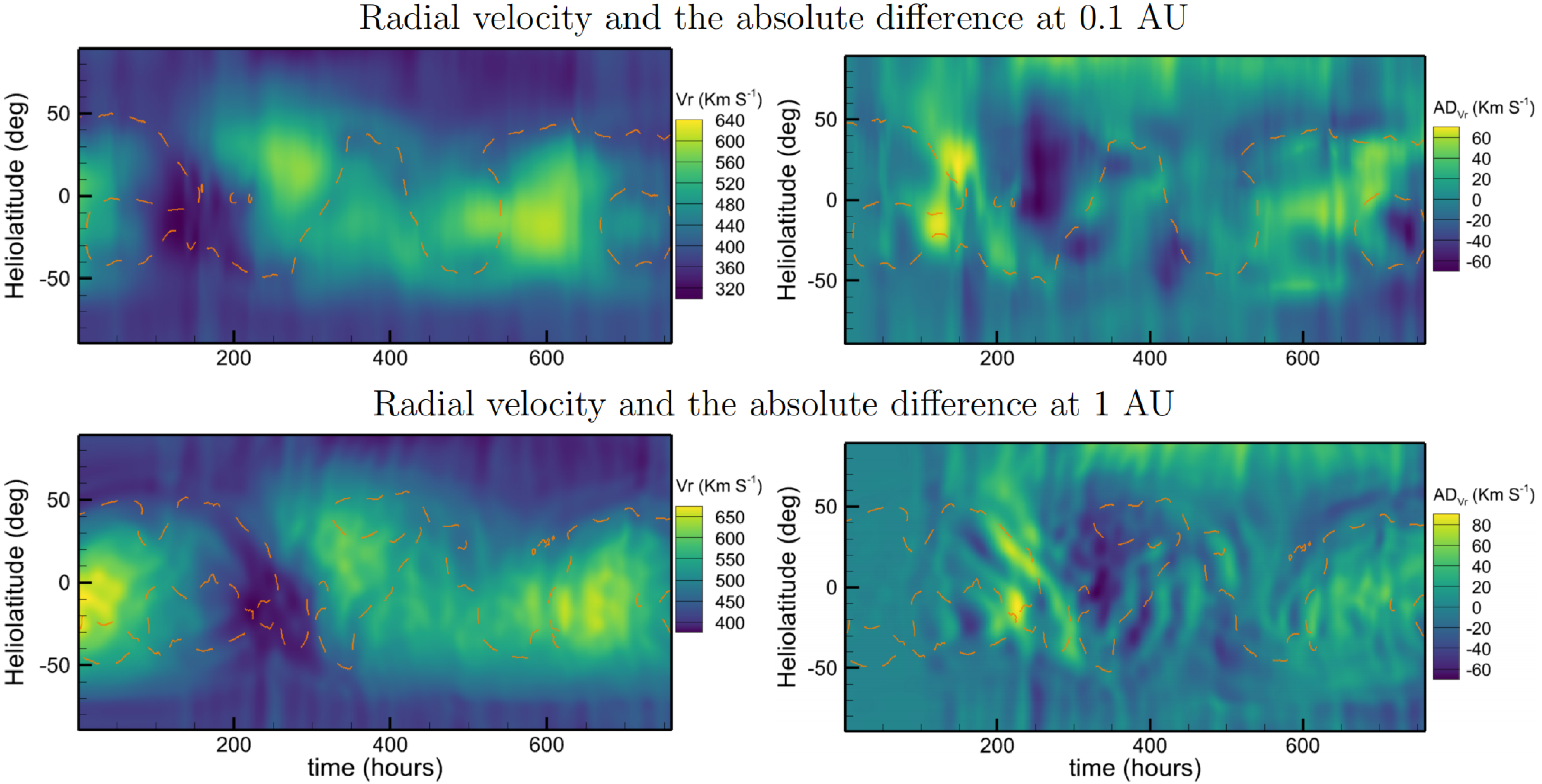}
\end{center}
\caption{Distributions of radial velocity $V_r \, (\mathrm{km~S^{-1}}; \, \rm left)$ along the longitude intersecting the Sun-Earth line, calculated from the time-evolving Sun-to-Earth simulation, and the corresponding absolute differences ${\rm AD}_{V_r} \, (\mathrm{km~S^{-1}}; \, \rm right)$ between the time-evolving and quasi-steady-state simulations, presented at 0.1 AU (top) and 1 AU (bottom). The overlaid orange dashed lines indicate the magnetic field neutral lines (MNLs) derived from the time-evolving simulation.}\label{RadialvelocityandAD}
\end{figure*}
The temporal evolution of the radial velocity distribution along the longitude intersecting the Sun-Earth line at 0.1~ AU (top left) and 1~AU (bottom left) is presented in Fig.~\ref{RadialvelocityandAD}.
Together with the distribution of magnetic field neutral lines (MNLs, orange dashed lines), they reveal that MNLs in the low- and middle-latitudes at both distances are consistently associated with low-speed flows, especially at 0.1~AU.
The velocity peaks around ($\theta,~t$)=($10^{\circ}$, 280~hours) and ($-20^{\circ}$, 600~hours), and the trough around ($\theta,~t$)=($-20^{\circ}$, 130~hours) at 0.1~AU, appear about 80 hours earlier than at 1~AU, and the temporal evolution of the radial velocity distributed along the selected longitude at 1~AU is consistent with that at 0.1~AU when an 80-hour time lag is considered. This suggests that the large-scale solar wind streamer structures are largely preserved during their propagation from 0.1 AU to 1 AU, meanwhile the Sun rotates by about $50^{\circ}$.

Figure~\ref{RadialvelocityandAD} also presents the absolute differences in radial velocity between the time-evolving and quasi-steady simulations, denoted by ${\rm{AD}_{Vr}}={\rm V_r^{QSS}-V_r^{TE}}$ with the superscripts $^{\rm QSS}$ and $^{\rm TE}$ denoting the corresponding variables calculated in the quasi-steady-state and time-evolving simulations. It reveals that ${\rm{AD}_{Vr}}$ is relatively small during the first 7 hours at 0.1~AU and during the first 90 hours at 1~AU, when perturbations caused by the evolving inner-boundary magnetic field have not yet reached the selected region, compared with the following times. In the subsequent period, the differences become pronounced, highlighting the significant impact of magnetogram evolution, often neglected in coronal and solar wind simulations for computational simplicity, on solar wind structures. 

\begin{figure*}[htpb]
\begin{center}
  \vspace*{0.01\textwidth}
    \includegraphics[width=0.8\linewidth,trim=1 1 1 1, clip]{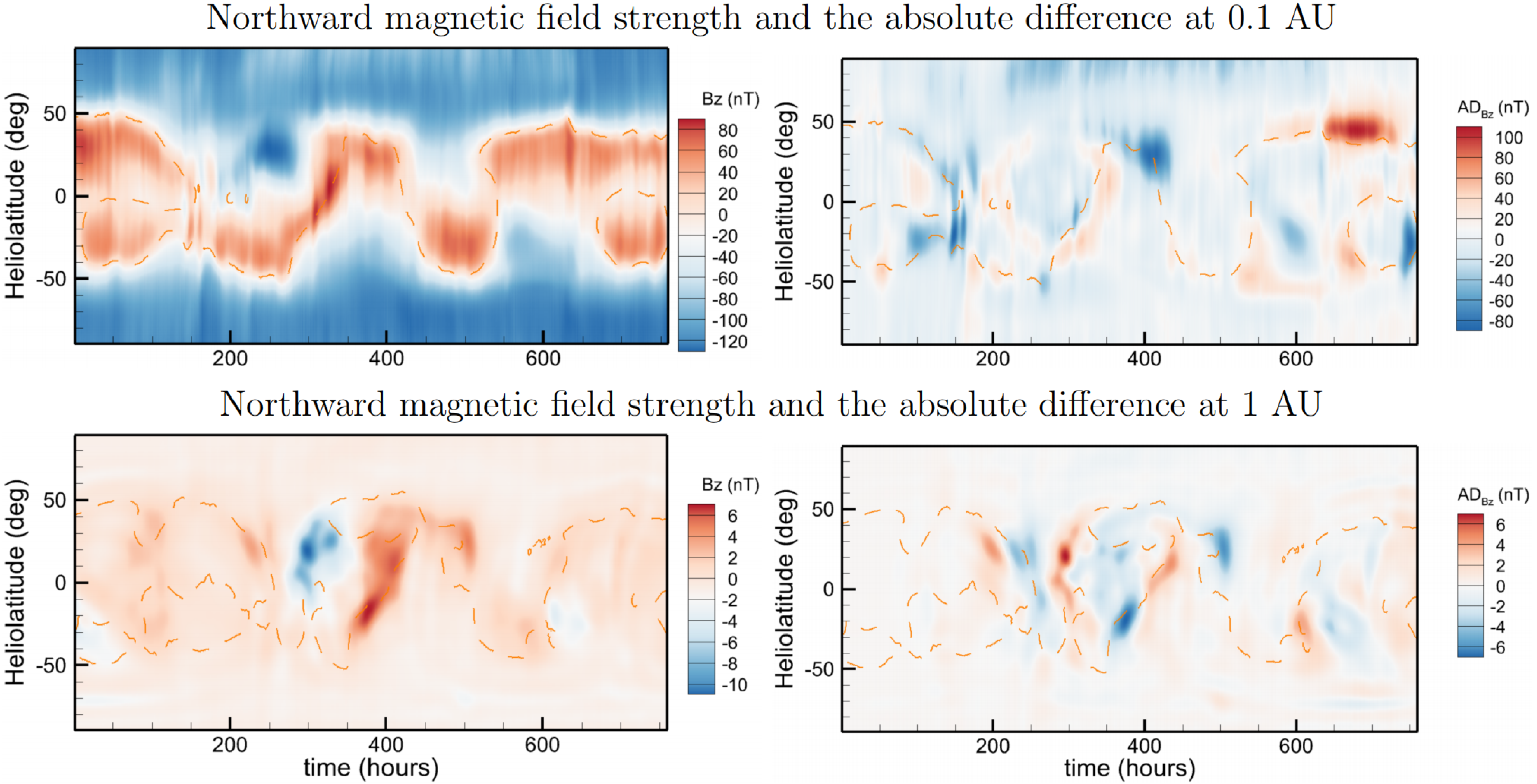}
\end{center}
\caption{Distributions of the northward magnetic field strength ${\rm B}_z \, (\mathrm{nT};\, \mathrm{left})$ along the longitude intersecting the Sun-Earth line, calculated from the time-evolving Sun-to-Earth simulation, and the corresponding absolute differences ${\rm AD}_{{\rm B}_z} \, (\mathrm{nT};\, \mathrm{right})$ between the time-evolving and quasi-steady-state simulations, presented at 0.1~AU (top) and 1~AU (bottom). The overlaid orange dashed lines indicate the magnetic neutral lines (MNLs) derived from the time-evolving simulation.}\label{BzandADBz}
\end{figure*}
Fig.~\ref{BzandADBz} shows the distribution of the northward magnetic field strength (left), as well as the absolute differences ${\rm{AD}_{Bz}}={\rm B_z^{QSS}-B_z^{TE}}$ between the time-evolving and quasi-steady simulations (right), at 0.1~AU (top) and 1~AU (bottom). It indicates that $\rm B_z$ at 0.1~AU is clearly non-uniform across latitude at 0.1 AU, but becomes nearly uniform along latitude in most regions by 1 AU. This is consistent with \textit{Ulysses} \citep{Wenzel1992,Smith1995} observations, which indicate that the magnetic field is nearly uniform with latitude at a given heliocentric distance \citep[except near the heliospheric current sheet;][]{Balogh1995,Smith1995,Lockwood2004}, as well as with MHD simulation results suggesting that the magnetic field may continue to vary with latitude out to at least 10~$R_s$ \citep{Reville_2017}. It also demonstrates that the absolute differences in the northward magnetic field strength between the time-evolving and quasi-steady-state simulations are significant at both 0.1 and 1 AU, particularly near the MNLs, with maximum magnitudes comparable to the maximum northward magnetic field strength.

\begin{figure*}[htpb]
\begin{center}
  \vspace*{0.01\textwidth}
    \includegraphics[width=0.8\linewidth,trim=1 1 1 1, clip]{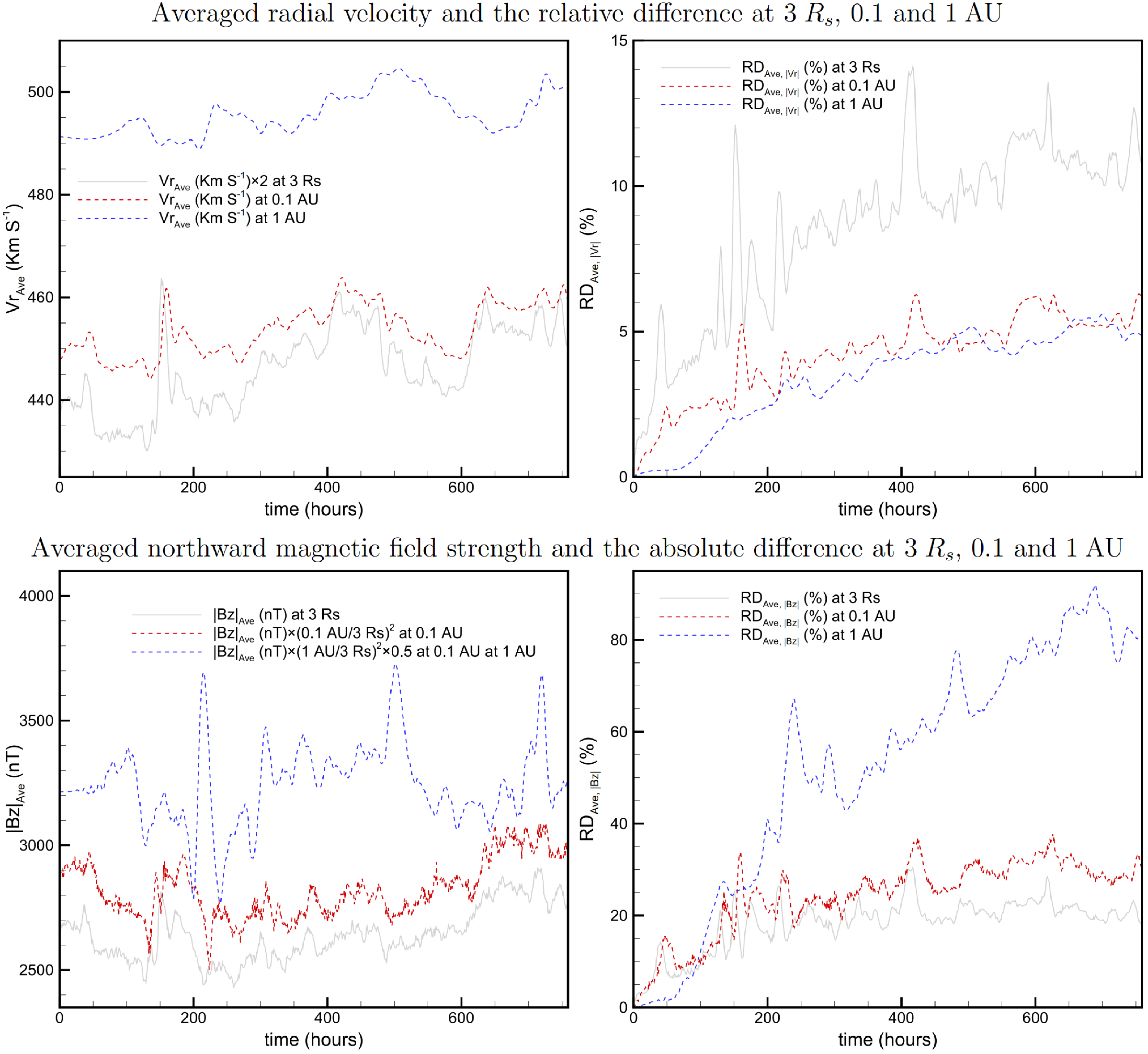}
\end{center}
\caption{Timing diagrams of the averaged radial velocity ${\rm V}_{r,\rm Ave}\,(\mathrm{km~s^{-1}};\,\mathrm{top~left})$ and the corresponding relative differences ${\rm RD}_{\rm Ave, V_r}\,(\mathrm{km~s^{-1}};\,\mathrm{top~right})$ between the time-evolving and quasi-steady-state simulations, evaluated at $3~R_s$ (grey solid lines), $0.1~\mathrm{AU}$ (red dashed lines), and $1~\mathrm{AU}$ (blue dashed lines). 
Also shown are the timing diagrams of the averaged northward magnetic field strength $|{\rm B_z}|_{\rm Ave}\,(\mathrm{nT};\,\mathrm{bottom~left})$ and the corresponding relative differences ${\rm RD}_{\rm Ave, |B_z|}\,(\mathrm{nT};\,\mathrm{bottom~right})$.}\label{VrandBzandAveragedRDVrandBz}
\end{figure*}
Fig.~\ref{VrandBzandAveragedRDVrandBz} further illustrates the averaged radial velocity ${\rm Vr}_{\rm Ave}$ (top left) and the averaged northward magnetic field strength (bottom left), along with the corresponding averaged relative differences in radial velocity ${\rm RD}_{\rm Ave,|Vr|}$ (top right) and northward magnetic field strength ${\rm RD}_{\rm Ave,|Bz|}$ (bottom right) between the time-evolving and quasi-steady-state simulations at 3~$R_s$, 0.1~AU and 1~AU.
This indicates that the averaged radial velocity increases predominantly within 0.1~AU, from 220~$\mathrm{km~s^{-1}}$ at 3~$R_s$ to approximately 450~$\mathrm{km~s^{-1}}$ at 0.1~AU. Over the domain between 0.1 AU and 1 AU, the value increases by approximately 10$\%$.
In addition, the temporal evolution profile of the averaged radial velocity becomes smoother with increasing heliocentric distance.
Consequently, the averaged relative difference ${\rm RD}_{\rm Ave,|Vr|}$ decreases with increasing heliocentric distance. However, it can still reach 5$\%$ at 1~AU. Here, the averaged values are defined as:
\begin{equation*}
{\rm Vr}_{\rm Ave}=\frac{\sum\limits_{i=1}^{N_{r=R_{Se}}} {\rm Vr}_{i}\cdot S_i}{\sum\limits_{i=1}^{N_{r=R_{Se}}} S_i},~~~~{\rm RD}_{{\rm Ave,|Vr|}}=\frac{\sum\limits_{i=1}^{N_{r=R_{Se}}} \left|{\rm Vr}_{i}^{\rm QSS}-{\rm Vr}_{i}^{\rm TE}\right|\cdot S_i}{\sum\limits_{i=1}^{N_{r=R_{Se}}} {\rm |Vr|}_{i}^{\rm TE}\cdot S_i}\,,
\end{equation*}
where $N_{r=R_{Se}}$ denotes the number of grid cells at $r=R_{Se}$, and $S_i$ represents the area of the $i$th grid cell at this distance.

Meanwhile, the averaged scaled unsigned northward magnetic field strength at $r=R_{Se}$, denoted by $|{\rm B_z}|_{\rm Ave}\times\left(\frac{R_{Se}}{3~R_s}\right)^2$, increases by approximately 8$\%$ and 23$\%$ from 3~$R_s$ to 0.1~AU and 1~AU, respectively. This increase can be attributed to inversions of heliospheric magnetic field lines \citep{Erdos_2012,Erdos_2014,Lockwood2009,Owens2017JGR,Frost2022}, which account for approximately 20$\%$ of the measured magnetic flux in ACE data \citep{McComas1998,Smith1998,king2005JGR} collected during 1998 and 2011 \citep{Owens2017JGR}. In contrast to ${\rm RD}_{\rm Ave,|Vr|}$, the averaged relative difference in the northward magnetic field strength ${\rm RD}_{\rm Ave,|B_z|}$ increases with heliocentric distance.
This can be explained by the fact that, at larger heliocentric distances, the magnetic field no longer dominates the solar wind and is increasingly influenced by the evolution of plasma flows, in addition to variations in the inner-boundary magnetic field. 
In Fig.~\ref{VrandBzandAveragedRDVrandBz}, the averaged variable are calculated as follows:
\begin{equation*}
{\rm |B_z|}_{\rm Ave}=\frac{\sum\limits_{i=1}^{N_{r=R_{Se}}} {\rm |B_z|}_{i}\cdot S_i}{\sum\limits_{i=1}^{N_{r=R_{Se}}} S_i},~~~~{\rm RD}_{{\rm Ave,|B_z|}}=\frac{\sum\limits_{i=1}^{N_{r=R_{Se}}} \left|{\rm B_z}_{i}^{\rm QSS}-{\rm B_z}_{i}^{\rm TE}\right|\cdot S_i}{\sum\limits_{i=1}^{N_{r=R_{Se}}} {\rm |Bz|}_{i}^{\rm TE}\cdot S_i}\,.
\end{equation*}

\subsection{Performance of the time-evolving Sun-to-Earth COCONUT}\label{sec:performance}
In this subsection, we further compare the Sun-to-Earth COCONUT simulation results at the L1 and L5 points with those from the commonly used EUHFORIA IH simulation \citep{Pomoell2018020,Poedts_2020,Linan_2025}, which is driven by a sequence of four-hourly updated Sun-to-Earth COCONUT simulation results at 0.1~AU. 
We also compare the computational efficiency of the Sun-to-Earth COCONUT with different time-step lengths. 

In the EUHFORIA simulation, 256 points are adopted in the radial direction from 0.1 to 2 AU, with 4-degree resolution in longitude and colatitude, covering 30 to 150 degrees. The time-step size of EUHFORIA can go down to a few minutes, and the time interval between two consecutive COCONUT output files is typically 4 hours, with only a few instances of 2-hour intervals at the initial stage. Therefore, substantial temporal interpolation is implemented between adjacent COCONUT output files. 
The EUHFORIA simulation includes a 10-day relaxation phase used to establish the initial background solar wind, and 32 days of solar wind evolution following the relaxation phase. The EUHFORIA solver has been recently ported to GPU, which will be described in a future paper. It requires 1 hour and 10 minutes when conducted on 3 MPI processes on a single GPU node. Although the runtime is about 1 hour, a full EUHFORIA pipeline requires more than that: one needs a coronal model and its outputs. So the total time includes running COCONUT, generating the series of inner boundary files, and running EUHFORIA. A detailed description of the coupling between COCONUT and EUHFORIA is available in \cite{Linan_2025}. For simplicity in practical implementation, we are currently working on automating the coupling between COCONUT and EUHFORIA.

The main differences between EUHFORIA and COCONUT are that the heating source and radiative loss terms included in COCONUT are not considered in EUHFORIA, and that the adiabatic index in COCONUT is $\gamma=1.67$, whereas $\gamma=1.5$ in EUHFORIA. Additionally, EUHFORIA employs explicit temporal integration, excludes the polar regions that are included in COCONUT, and adopts a constrained transport approach \citep{Yee1138693,EvansandHawley1988} to ensure that the magnetic field remains divergence-free. Besides, the energy decomposition strategy, which significantly improves the numerical stability of COCONUT, has not yet been implemented in EUHFORIA.  Moreover, the hydrogen population $m_{\rm cor}$ assumption differs between COCONUT and EUHFORIA, with $m_{\rm cor}=1.27$ in COCONUT and $m_{\rm cor}=0.5$ in EUHFORIA.

\begin{figure*}[htpb]
\begin{center}
  \vspace*{0.01\textwidth}
    \includegraphics[width=\linewidth,trim=1 1 1 1, clip]{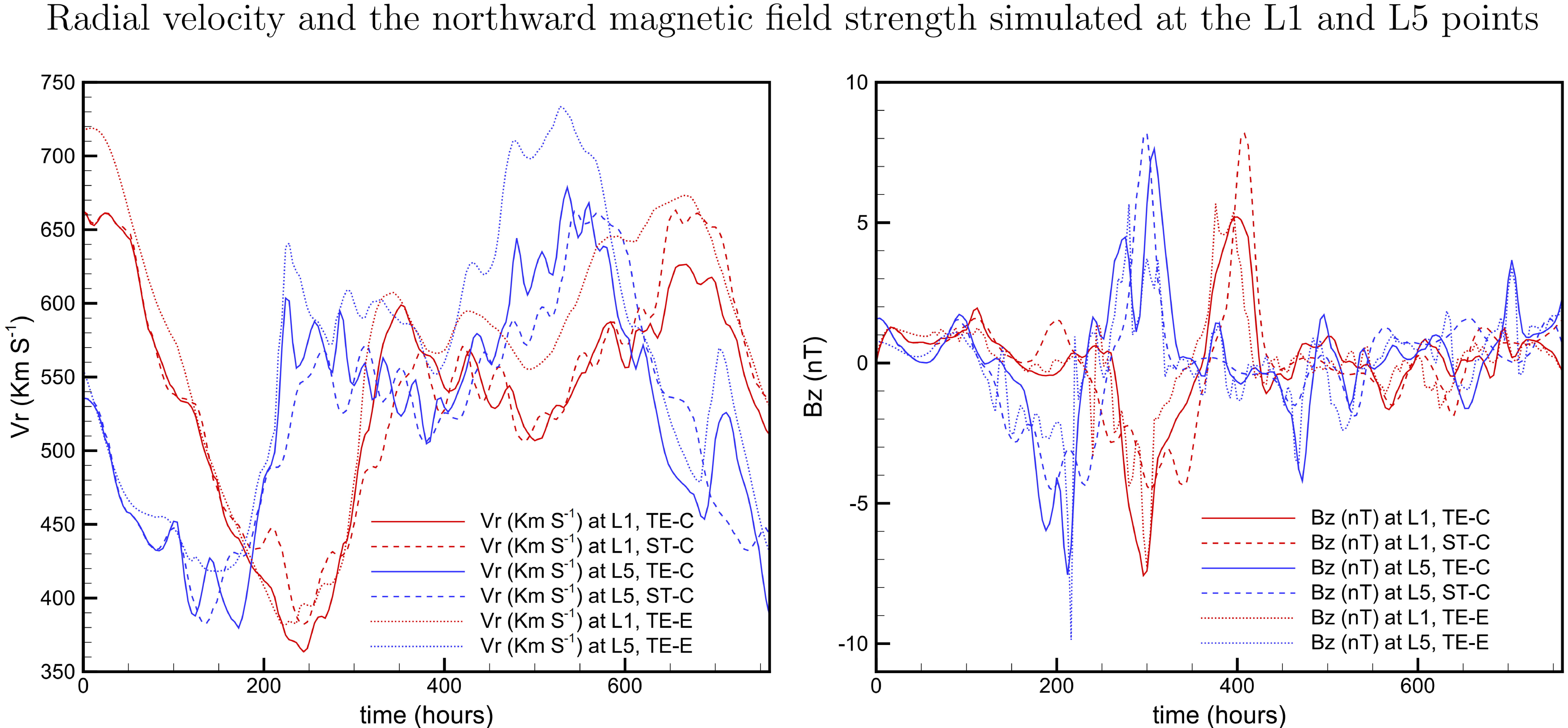}
\end{center}
\caption{Timing diagrams of the radial velocity $V_r \, (\mathrm{km~S^{-1}}; \, \rm left)$ and northward magnetic field strength ${\rm B}_z \, (\mathrm{nT};\, \mathrm{left})$ simulated by the time-evolving (TE; solid lines) and quasi-steady-state (ST; dashed lines) Sun-to-Earth COCONUT and observed by two virtual satellites placed at the L1 and L5 points during the simulated period. The dashed lines represent the corresponding EUHFORIA simulation results, driven by the time-evolving Sun-to-Earth COCONUT simulation results at 0.1 AU.}\label{L1andL5}
\end{figure*}
In Fig.~\ref{L1andL5}, we present the simulated timing diagrams of radial velocity and northward magnetic field strength at the L1 (red) and L5 (blue) points obtained from the time-evolving Sun-to-Earth COCONUT simulation, the time-evolving IH model EUHFORIA driven by a series of COCONUT results at 0.1~AU, and the quasi-steady-state Sun-to-Earth COCONUT simulation with longitude mapped to time following Eq.~(\ref{phitotime}).
For convenience, we denote the time-evolving and steady-state COCONUT simulations as ``TE-C'' and ``ST-C'', respectively, and refer to the time-evolving EUHFORIA simulation as ``TE-E''.
This indicates that the radial velocity and northward magnetic field at L1 and L5 in both the time-evolving COCONUT and EUHFORIA simulations are consistent with each other, except for the higher radial velocity in the EUHFORIA simulation between 230 and 580 hours at L1, and between 400 and 720 hours at L5. 

\begin{table*}
\centering
\caption{Carrington rotation-averaged radial velocity and northward magnetic field strength, and their relative differences, at L1 and L5 points.}
\label{QSSVSTE_Vr}
\begin{tabular}{lllll}
\hline\noalign{\smallskip}
 Variables & ${\rm Vr}_{\rm ave}^{\rm TE-C}~\& ~{\rm Vr}_{\rm ave}^{\rm TE-E}$  & ${\rm RD}_{\rm ave,\left|V_r\right|}^{\rm TE-C:TE-E}$~\&~${\rm RD}_{\rm ave,\left|V_r\right|}^{\rm ST-C:TE-C}$  & ${\rm |B_z|}_{\rm ave}^{\rm TE-C}~\&~{\rm |B_z|}_{\rm ave}^{\rm TE-E}$ & ${\rm RD}_{\rm ave,\left|B_z\right|}^{\rm TE-C:TE-E}$~\&~${\rm RD}_{\rm ave,\left|B_z\right|}^{\rm ST-C:TE-C}$ \\
\noalign{\smallskip}\hline\noalign{\smallskip}
 L1 point & 523.16~\&~554.71~$\mathrm{km~S^{-1}}$  & $6.16\%$~\&~$3.85\%$ & 1.24~\&~0.97~$\rm nT$ & $56.37\%$~\&~$70.29\%$ \\
 L5 point & 534.15~\&~573.30~$\mathrm{km~S^{-1}}$ & $7.39\%$ \& $5.18\%$ & 1.52~\&~1.23~$\rm nT$ & $62.75\%$ \& $92.39\%$\\
\noalign{\smallskip}\hline
\end{tabular}
\end{table*}
Figure~\ref{L1andL5} also shows that, unlike the quasi-steady-state simulation, where the temporal profiles at L1 and L5 point are identical except for a $\sim$110-hour lead at L5 point, the time-evolving simulation results at L1 and L5 points exhibit noticeable differences in both the shape and magnitude of the corresponding peaks and troughs in the radial velocity and northward magnetic field strength profiles.
In Table~\ref{QSSVSTE_Vr}, we further present the temporally averaged radial velocity and northward magnetic field strength at the L1 and L5 points over one CR, corresponding to 100 to 752 hours of the time-evolving simulation period, in the time-evolving COCONUT and EUHFORIA (${\rm Vr}_{\rm ave}^{\rm TE-E}$ and ${\rm |B_z|}_{\rm ave}^{\rm TE-E}$) simulations, together with the averaged relative differences between the time-evolving coconut and EUHFORIA simulations (${\rm RD}_{\rm ave,\left|V_r\right|}^{\rm TE-C:TE-E}$ and ${\rm RD}_{\rm ave,\left|B_z\right|}^{\rm TE-C:TE-E}$), and between the time-evolving and quasi-steady-state COCONUT simulations (${\rm RD}_{\rm ave,\left|V_r\right|}^{\rm ST-C:TE-C}$ and ${\rm RD}_{\rm ave,\left|B_z\right|}^{\rm ST-C:TE-C}$).

In Table~\ref{QSSVSTE_Vr}, the averaged relative differences are calculated as follows:
\begin{equation*}
\begin{aligned}
{\rm RD}_{\rm ave,\left|\xi\right|}^{\rm TE-C:TE-E}=&\frac{\sum\limits_{i=1}^{N_{\rm CR}}\left|\xi^{\rm TE-E}-\xi^{\rm TE-C}\right|}{\sum\limits_{i=1}^{N_{\rm CR}}\left|\xi^{\rm TE-C}\right|}\\
{\rm RD}_{\rm ave,\left|\xi\right|}^{\rm ST-C:TE-C}=&\frac{\sum\limits_{i=1}^{N_{\rm CR}}\left|\xi^{\rm TE-C}-\xi^{\rm ST-C}\right|}{\sum\limits_{i=1}^{N_{\rm CR}}\left|\xi^{\rm ST-C}\right|}
\end{aligned}\,, ~~~\xi \in (\rm V_r,~B_z)\,.
\end{equation*} 
Here, $N_{\rm CR}$ denotes the number of sample points adopted during one CR at L1 or L5 point, and we adopt $N_{\rm CR}=163$ in this paper. 
This shows that the differences between the time-evolving COCONUT and EUHFORIA simulations are comparable to those between the time-evolving and quasi-steady-state COCONUT simulations and that 
EUHFORIA yields higher radial velocities and lower northward magnetic field strengths than COCONUT, although the results are broadly consistent. 
The discrepancies can be attributed to differences in the governing equations and numerical algorithms between COCONUT and EUHFORIA. In addition, the spatial interpolation required to construct the interface at 0.1~AU for EUHFORIA from the COCONUT mesh, which does not coincide with the EUHFORIA grid, as well as the temporal interpolation from input files to the EUHFORIA time steps, also introduces differences. In summary, both spatial and temporal interpolations affect the results, further motivating the use of a single model capable of handling the full evolution.

\section{Concluding remarks}\label{sec:Conclusion}

Limited by the Courant-Friedrichs-Lewy numerical stability constriction, most currently used coronal MHD models require much smaller time steps than IH models. Therefore, Sun-to-Earth modelling frameworks typically couple separate coronal and IH models at a prescribed interface to enable practical applications. However, differences in numerical schemes, grid structures, boundary prescriptions, and physical assumptions between the two models often introduce uncertainties. Additionally, coupling two models adds extra complexity, making the Sun-to-Earth modelling pipeline more cumbersome. Developing implementation-ready Sun-to-Earth solar corona and solar wind MHD models that combine efficiency, accuracy, and numerical stability is crucial for practical applications such as daily space weather forecasting.

Given that the implicit time integration method with Newton iterations enables the time-evolving coronal MHD model COCONUT to achieve high computational efficiency while maintaining the desired temporal accuracy \citep{Wang2025_FirsttimeevolvingCOCONUT,wang2025COCONUTMayEvent}, and that the energy decomposition strategy significantly improves its numerical stability \citep{Wang_COCONUT_DecE}, we further extend this model to 1 AU in this paper. This demonstrates that a single MHD model such as COCONUT, which adopts an advanced implicit scheme, can cover the entire Sun-to-Earth domain while effectively balancing computational efficiency, numerical stability, and modelling accuracy.
It thereby eliminates the need for the multi-step process of first running a coronal model and then driving an IH model using pre-saved coronal outputs, and avoids the uncertainties associated with coupling two distinct models.

Using this efficient time-evolving Sun-to-Earth COCONUT MHD model, we simulate the continuous evolution of solar corona and wind from 1.01~$R_s$ to beyond Earth's orbit over one solar maximum CR to investigate this Sun-to-Earth modelling approach. The simulation reproduces the latitudinal invariance of the magnetic field distribution at 1~AU derived from \textit{Ulysses} observations \citep{Wenzel1992,Smith1995}. The time-evolving Sun-to-Earth COCONUT simulation results at the L1 and L5 points are also consistent with those from the commonly used IH model EUHFORIA which is driven by the corresponding COCONUT simulation results at 0.1~AU. The peaks and troughs in the timing diagrams of variables at the L5 point occur about four days earlier than those at L1, demonstrating that L5 observations can be used to predict solar wind conditions near Earth approximately four days in advance. Meanwhile, obvious differences in the magnitude and shape of these peaks and troughs highlight the necessity of conducting multi-satellite observations for reconstructing a more realistic, synchronised global ($4\pi$) solar corona and wind. 

Additionally, a comparison between the time-evolving and quasi-steady-state Sun-to-Earth COCONUT simulations demonstrates that the impact of magnetogram evolution during a solar maximum CR on the simulated solar coronal and wind structures is underestimated in commonly used quasi-steady-state simulations, which assume that solar coronal and wind evolution over one CR is negligible. 
Besides, a comparison between the simulation results of the time-evolving Sun-to-Earth and coronal COCONUT models shows small but non-zero differences, which are mainly concentrated in narrow regions around the MNLs.
It indicates that an oversimplified outer boundary treatment, such as purely radial interpolation in 3D cases and zero-gradient interpolation for imposing outer boundary conditions of non-linear solar wind MHD system, can still introduce numerical artifacts and affect the solution within the inner computational domain, even when the solar wind is already supersonic and super-Alfv{\'e}nic at the outer boundary. 

Although the time-evolving Sun-to-Earth MHD model COCONUT achieves an effective balance between computational efficiency, numerical stability, and modeling accuracy, several challenges remain and  need to be addressed to further improve the current time-evolving Sun-to-Earth MHD model COCONUT. The currently adopted static grid mesh cannot capture transient events such as CMEs with sufficient fidelity throughout their propagation across the entire computational domain. The empirically defined heating source term is overly simplified and have difficulty to accurately reproduce the true heating rate. The solar atmosphere between the photosphere and the low corona is not yet included, and interpolating photospheric magnetic field observations to the low corona, around 1.01~$R_s$, to define the inner boundary magnetic field conditions introduces non-negligible uncertainty.

The ultimate objective of coronal-heliospheric simulation in numerical space weather modelling is to reproduce ambient solar wind of both magnetic structures and background flow characteristics, and solar eruptive events/their propagation in the ambient solar wind obtained, by using first-principles physics-based model driven and constrained solely by observations. The solar wind is a vital component of space weather, providing a background for solar transients such as CMEs, stream interaction regions, and energetic particles propagating toward Earth. Accurate prediction of space weather events requires a precise description and thorough understanding of physical processes occurring in the solar eruptive region and ambient solar wind plasma.  Achieving this would minimize reliance on theoretical assumptions and empirical relations. However, due to gaps in our physical understanding, incomplete observational data, and limited computational resources, fully realising this goal remains a challenge. 

In future work, based on the faster-than-real-time, time-evolving Sun-to-Earth MHD model COCONUT presented in this paper, we will advance to simulating CME propagation within a time-evolving solar wind background rather than the commonly used quasi-static background. Furthermore, we will incorporate the transition region \citep{Lionello_2009,Mok_2005,Mikic_2013,MIKIC2018NatA,Downs2025PSI}, include more realistic and physically consistent heating and acceleration mechanisms \citep{Cranmer_2010,Schleich2023,VanDoorsselaere2025}, and aim to perform faster-than-real-time, self-consistent CME initiation and eruption simulations driven by continuously evolving high-cadence and high-resolution active-region magnetograms \citep{Jiang2021fRONTIER,Jiang2023}. For practical applications, we will also implement a local time-stepping approach \citep{Groth2000,TOTH2012870} in the time-evolving Sun-to-Earth MHD model COCONUT once it has incorporated high-resolution active regions. Meanwhile, we will explore the use of characteristic boundary conditions \citep{Hayashi_2021,Feng_2023,Tarr_2024,Kee_2025} to mitigate numerical artifacts.


\begin{acknowledgements}
This project has received funding from the European Research Council Executive Agency (ERCEA) under the ERC-AdG agreement No. 101141362 (Open SESAME). 
These results were also obtained in the framework of the projects FA9550-18-1-0093 (AFOSR), C16/24/010  (C1 project Internal Funds KU Leuven), G0B5823N and G002523N (WEAVE) (FWO-Vlaanderen), and 4000145223 (SIDC Data Exploitation (SIDEX), ESA Prodex).
This work is also supported by the National Natural Science Foundation of China (grant No. 42030204) and the BK21 FOUR program of the Graduate School, Kyung Hee University (GSX-20242364 and GSX-20253142). The Research Council of Norway supports FZ through its Centres of Excellence scheme, project number 262622.
The resources and
services used in this work were provided by the VSC (Flemish Supercomputer Centre), funded by the Research Foundation – Flanders (FWO) and the Flemish Government. This work utilises data obtained by the Global Oscillation Network Group (GONG) program, managed by the National Solar Observatory and operated by AURA, Inc., under a cooperative agreement with the National Science Foundation. The data were acquired by instruments operated by the Big Bear Solar Observatory, High Altitude Observatory, Learmonth Solar Observatory, Udaipur Solar Observatory, Instituto de Astrof{\'i}sica de Canarias, and Cerro Tololo Inter-American Observatory. The authors also acknowledge the use of the STEREO/SECCHI data produced by a consortium of the NRL (US), LMSAL (US), NASA/GSFC (US), RAL (UK), UBHAM (UK), MPS (Germany), CSL (Belgium), IOTA (France), and IAS (France). 
\end{acknowledgements}

\bibliographystyle{aa}
\bibliography{SIPandCOCONUT}

\begin{appendix}
\section{Influence of the inner heliosphere  component on coronal simulation results}\label{IHimpactonCoronal}
To evaluate the impact of IH evolution on coronal simulation results, we performed a time-evolving coronal COCONUT simulation. In this appendix, we compare the results of the time-evolving coronal COCONUT simulation with those of the time-evolving Sun-to-Earth COCONUT simulation. 

\begin{figure}[htpb]
\begin{center}
  \vspace*{0.01\textwidth}
    \includegraphics[width=\linewidth,trim=1 1 1 1, clip]{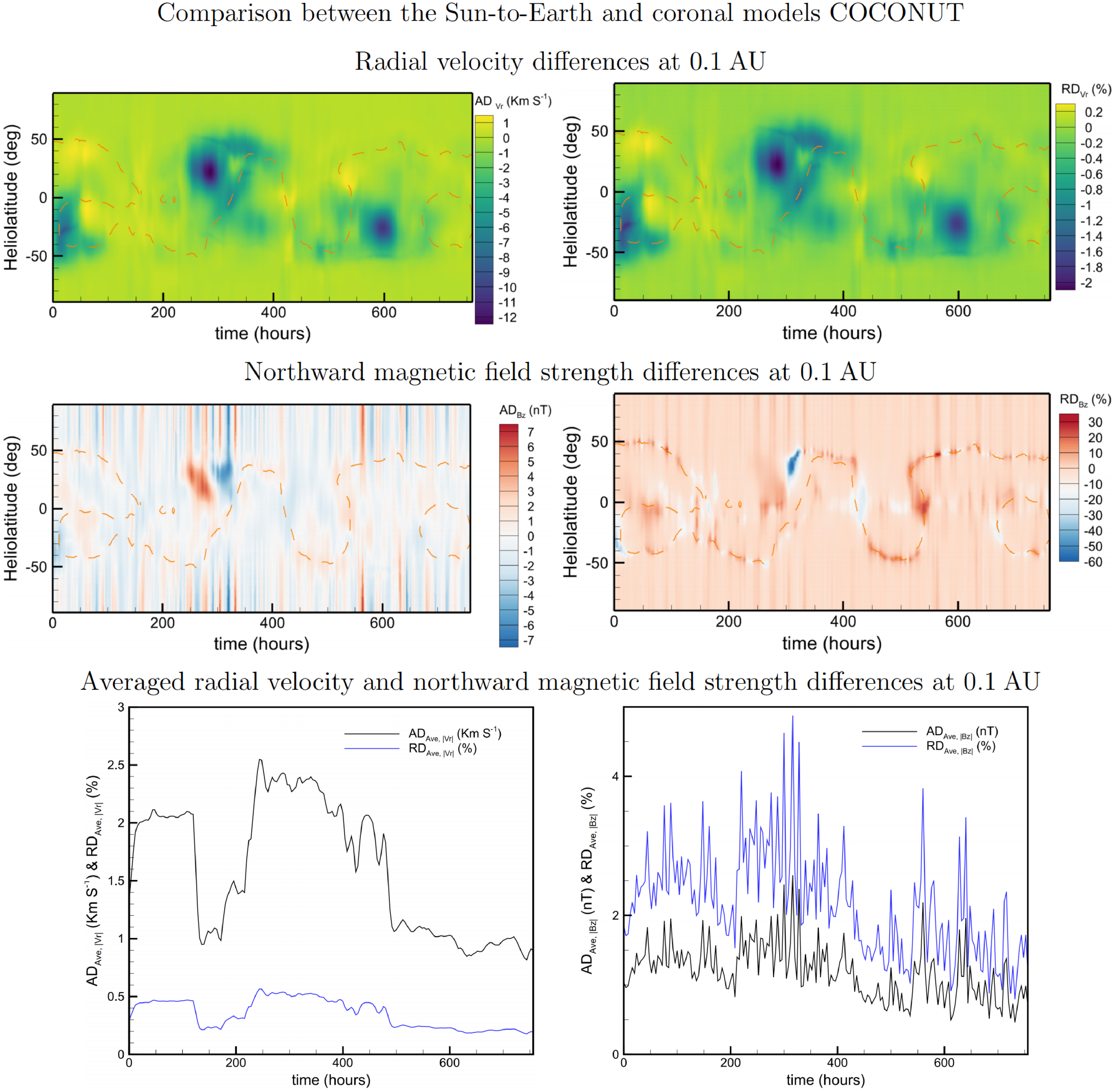}
\end{center}
\renewcommand{\thefigure}{A.~1}
\caption{Timing diagrams of the absolute and relative differences in radial velocity along the longitude intersecting the Sun--Earth line at 0.1~AU, comparing the Sun-to-Earth and coronal COCONUT simulations. These are denoted by ${\rm AD}_{\rm Vr}\,(\mathrm{km~s^{-1}})$ (top left) and ${\rm RD}_{\rm Vr}\,(\%)$ (top right), respectively. 
${\rm AD}_{\rm Bz}\,(\mathrm{nT})$ (middle left) and ${\rm RD}_{\rm Bz}\,(\%)$ (middle right) represent the corresponding absolute and relative differences in the northward magnetic field strength. 
The bottom panels show averaged quantities: ${\rm AD}_{\rm Ave,|Vr|}\,(\mathrm{km~s^{-1}})$ and ${\rm RD}_{\rm Ave,|Vr|}\,(\%)$ (bottom left) denote the averaged absolute and relative differences in radial velocity, while ${\rm AD}_{\rm Ave,|Bz|}\,(\mathrm{nT})$ and ${\rm RD}_{\rm Ave,|Bz|}\,(\%)$ (bottom right) denote the averaged absolute and relative differences in the magnetic field strength. 
All quantities are evaluated at 0.1~AU. 
}\label{SuntoEarthVScoronaat0d1AU}
\end{figure}

\begin{figure}[htpb]
\begin{center}
  \vspace*{0.01\textwidth}
    \includegraphics[width=\linewidth,trim=1 1 1 1, clip]{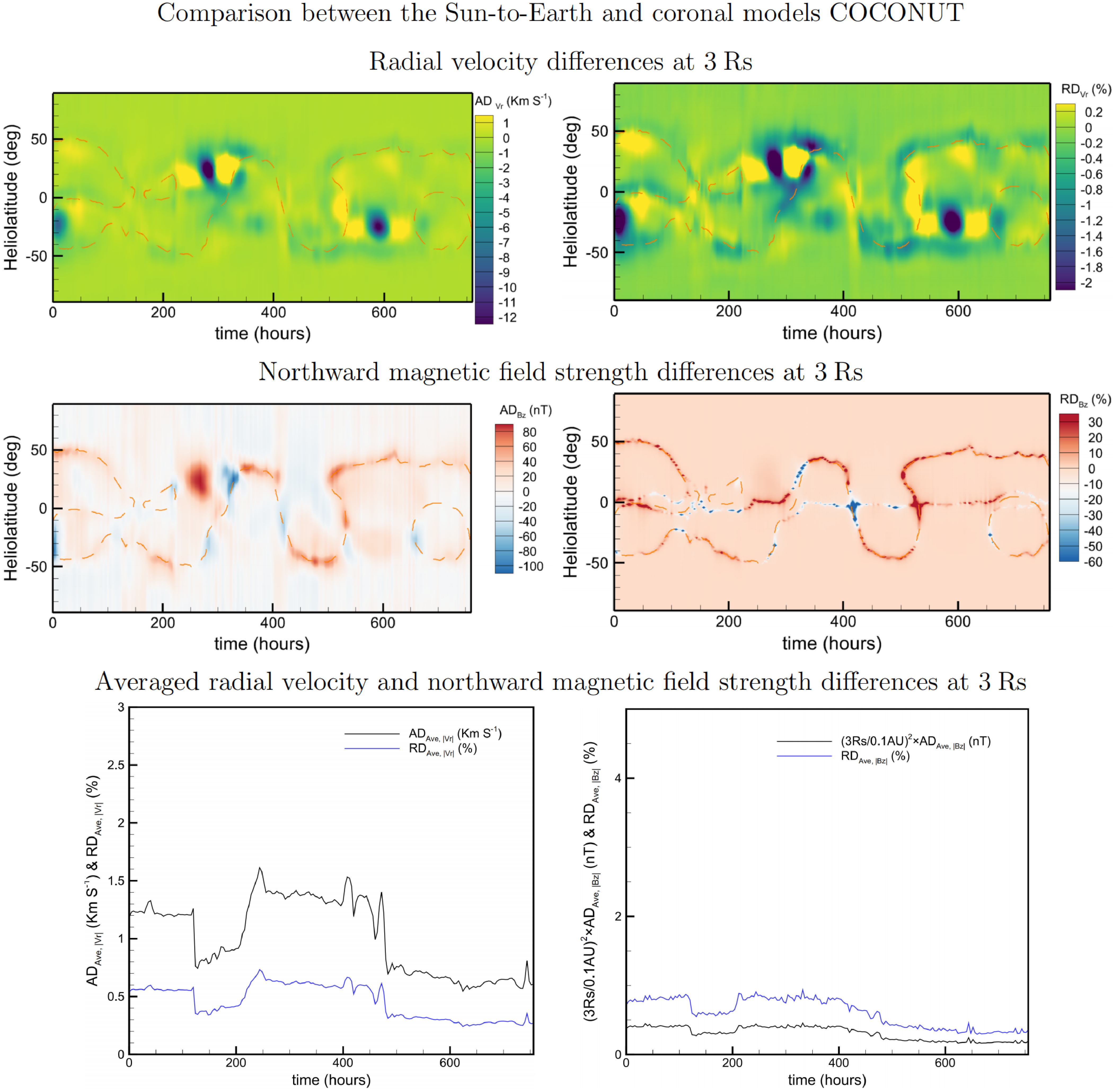}
\end{center}
\renewcommand{\thefigure}{A.~2}
\caption{The same as Fig.~\ref{SuntoEarthVScoronaat0d1AU}, but evaluated at 3~$R_s$}\label{SuntoEarthVScoronaat3Rs}
\end{figure}

Figures~\ref{SuntoEarthVScoronaat0d1AU} and \ref{SuntoEarthVScoronaat3Rs} reveal that, although the differences are very small compared to those between the time-evolving and quasi-steady-state simulation results, the inclusion of the IH component still affects the simulated coronal results. Additionally, both the absolute differences in radial velocity and scaled northward magnetic field strength, as well as the relative differences in the northward magnetic field strength, are smaller at 3~$R_s$ than at 0.1~AU, whereas the relative differences in radial velocity are similar at 3~$R_s$ and 0.1~AU. It is noticed that the averaged relative and absolute differences in radial velocity are smaller than 1$\%$ and 2.5 $\mathrm{km~s^{-1}}$ , respectively, during the entire simulation period at both 0.1~AU and 3~$R_s$. It is also observed that the averaged relative difference in the northward magnetic field strength increases from below 1$\%$ at 3~$R_s$ to approach 4$\%$ at 0.1~AU, with regions near the MNLs contributing most to the relatively large averaged difference. 

Considering that the solar wind is already supersonic and super-Alfv{\'e}nic at the 0.1~AU, setting the out boundary at 25~$R_S$ (the outer boundary of coronal model COCONUT) or 256~$R_s$ (the outer boundary of Sun-to-Earth model COCONUT) should not affect the  simulation results within 0.1~AU therapeutically. However, the solar wind does not propagate strictly radially at the boundaries of either the coronal or Sun-to-Earth models, implying that the currently adopted purely radial interpolation is insufficient for defining the outer boundary conditions without introducing numerical artifacts. Additionally, the differences, although very small and mainly concentrated in narrow regions around the MNLs, can also result from the currently implemented, oversimplified zero-gradient interpolation from the outermost cell centers in the computational domain to the ghost cells at the outer boundary \citep{Brchnelova2022,Brchnelova_2022,Perri_2022,Wang2025_FirsttimeevolvingCOCONUT}, which assumes a constant velocity beyond the outer boundary that is not consistent with the real conditions. 
We will adopt more accurate characteristic boundary conditions \citep{Feng_2012,Feng_2023,Hayashi_2021,Tarr_2024,Kee_2025}, which is reported to help prevent artificial X-points near the Sun that can arise in non-characteristic approaches \citep{Yalim_2017}, to mitigate the numerical influence of the boundary conditions on coronal evolution in future research work.

\end{appendix}

\end{document}